\documentclass[conference]{IEEEtran}
\usepackage{amsmath,amstext,amsfonts,latexsym,amssymb,graphicx}
\newcommand{\beq}{\begin{equation*}}
\newcommand{\eeq}{\end{equation*}}

\newcommand{\argmax}[1]{\arg{\hbox{$\underset{#1}{\max}\,$}}}

\newcommand{\beqq}{\begin{equation}}

\newcommand{\bi}{\begin{itemize}}

\newcommand{\eeqq}{\end{equation}}

\newcommand{\ei}{\end{itemize}}

\begin{document}
\title{A hybrid decision approach for the association problem in heterogeneous networks}

\author{\IEEEauthorblockN{Salah Eddine Elayoubi}
\IEEEauthorblockA{Orange Labs\\
38-40 Rue du General Leclerc\\
92130 Issy-Les-Moulineaux, France\\
salaheddine.elayoubi@orange-ftgroup.com}
\and
\IEEEauthorblockN{Eitan Altman}
\IEEEauthorblockA{INRIA Sophia Antipolis\\
10 route des Lucioles\\
06902 Sophia Antipolis, France\\
Eitan.Altman@sophia.inria.fr }
\and
\IEEEauthorblockN{Majed Haddad, Zwi Altman}
\IEEEauthorblockA{Orange Labs\\
38-40 Rue du General Leclerc\\
92130 Issy-Les-Moulineaux, France\\
\{majed.haddad,zwi.altman\}@orange-ftgroup.com}}
\maketitle
\begin{abstract}
The area of networking games has had a growing impact on wireless
networks. This reflects the recognition in the important scaling
advantages that the service providers can benefit from by increasing
the autonomy of mobiles in decision making. This may however result
in inefficiencies that are inherent to equilibria in non-cooperative games. Due to the
concern for efficiency, centralized
protocols keep being considered and compared to decentralized
ones. From the point of view of the network architecture, this
implies the co-existence of network-centric and terminal centric
radio resource management schemes. Instead of taking part within the
debate among the supporters of each solution, we propose in this
paper hybrid schemes where the wireless users are assisted in their
decisions by the network that broadcasts aggregated load
information.  We derive the utilities related to the
Quality of Service (QoS) perceived by the users and develop a
Bayesian framework to obtain the equilibria.  Numerical results
illustrate the advantages of using our hybrid game framework in an
association problem in a network composed of HSDPA and 3G LTE
systems.
\end{abstract}

\section{Introduction}
In order to handle the growing wireless traffic demand, operators
are often faced with the need to install new base stations. This could result in splitting cells into smaller
ones, or in having several base stations covering the same cell. The
second option may be preferred when the traffic has high variability
(in time and space) in which case it may be advantageous to have the
possibility to allocate resources from both base stations to any
point in the cell. This flexibility comes at a cost of having to
include an access control that takes the proper association
decisions for the mobiles, that of deciding to which base station
(BS) to connect. To achieve efficient use of the resources, these
decisions should be based not only on the current system state but
also on expected future demand which may interact
with traffic assigned in the present.

We wish to avoid completely decentralized solutions of the
association problem in which all decisions are taken by the mobiles,
due to well known inefficiency problems that may arise when each
mobile is allowed to optimize its own utility. This inefficiency is
inherent to the non-cooperative nature of the decision making. On
the other hand, we wish to delegate to the mobiles a large part in
the decision making in order to alleviate the burden from the base
stations.

The association schemes actually implemented
are fully centralized: the operator tries to maximize his utility
(revenue)  by assigning the users to the different systems
\cite{assocC1}-\cite{assocC3}. However, distributed RRM mechanisms
are gaining in importance: Users may be allowed to make autonomous
decisions in a distributed way. This has lead to game theoretic
approaches to the association problems in wireless networks, as can
be found in \cite{assocG1}-\cite{assocG5}. The
potential inefficiency of such approaches have been known for a long
time. The term "The Tragedy of the Commons" has been frequently used
for this inefficiency \cite{tragedy}; it describes a dilemma in
which multiple individuals acting independently in their own
self-interest can ultimately destroy a shared limited resource even
when it is clear that it is not in anyone's long term interest for
this to happen.

We propose in this paper association methods that combine benefits
from both decentralized and centralized design. Central intervention
is needed during severe congestion periods. At those instants, we
assume that the mobiles follow the instructions of the base
stations. Otherwise the association decision is left to the mobiles,
who make the decision based on aggregated state information from the
base stations. The decision making is thus based on partial
information that is signaled to the mobiles by the base station. A
central design aspect is then for the base stations to decide how to
aggregate information which then determines what to signal to the
users. Note that this decision making at the BS can  be viewed as a mechanism
design problem, or as a Bayesian game.

\section{Problem statement}

 \subsection{System description}

\begin{figure}[ht]
\vspace{-0.5cm}
\hspace{-0.5cm}
\includegraphics[width=10cm]{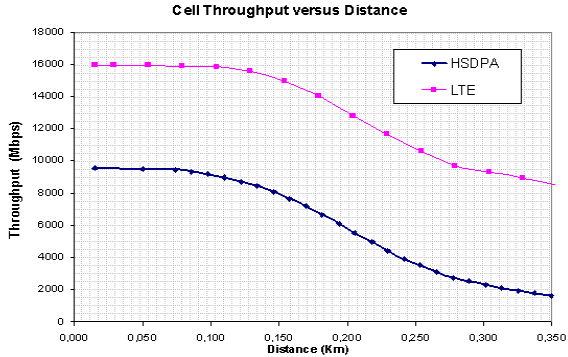}
\vspace{-1cm}
\caption{LTE and HSDPA peak throughputs for different user positions.} \label{fig:debit}
\end{figure}

We consider a network composed of $S$ systems operated
by the same operator. Even if the model we develop is applicable to different kinds of situations, we will focus
on the more realistic and cost effective case where the operator uses the same cell
sites to deploy the new system (e.g. 3G LTE), while keeping the old
ones (e.g. HSDPA). As, in each cell, there are different radio
conditions following the position of the user regarding the cell
site, the peak throughput that can be obtained by the user connected
to system $s$, if served alone by a cell, differs following his
position in the cell, as illustrated in Figure \ref{fig:debit} for a cell served by HSDPA and 3G LTE.
For simplicity, we consider that there are $N$ classes of radio
conditions and that users with radio condition $n$ have a peak rate
$D_n^s$ if connected to system $s$. The network state
is then defined by the vector: ${\bf M} =
(M_1^1,...,M_N^1,..., M_1^S,..., M_N^S)$, $M_n^s$ being
the number of users with radio condition $n$ connected to
system $s$.

\begin{figure}[ht]
\vspace{-0.5cm}
\centering
\includegraphics[width=8cm]{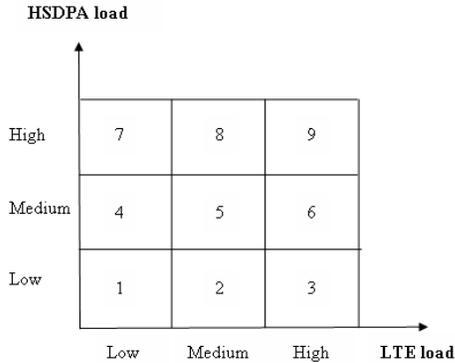}
\vspace{-1cm}
\caption{Aggregated load information.} \label{fig:pload}
\end{figure}

We assume that the network broadcasts a partial
load information $l$ ($1\leq l\leq L$), e.g., an aggregated load
information indicating for each system
if it is in low, medium, or
high load state. An example of this load information is described in
Figure \ref{fig:pload} for a network composed of HSDPA and LTE systems.

\subsection{Policy definition}
As stated before, users are only aware of the load information $l$ sent by the network. Their policies are then based on this information. Let ${\bf P}$ be a policy defined by the actions taken by mobiles in the different load conditions. ${\bf P}$ is a $N\times L$ matrix whose element $P_{nl}$ is equal to $s$ if class-$n$ users connect to system $s$ when the network broadcasts information $l$.

Let $\mathcal{A}$ be the space of feasible states, $\mathfrak{P}$ be the set of all possible policies and let $\mathfrak{L}$ be the set of
load information. An assignment $f : \displaystyle \mathcal{A}\rightarrow \mathfrak{L}$ specifies for each network state ${\bf M}$ the
corresponding load information $f({\bf M})$. On the other hand, when the load information is equal to $l$ and the policy is ${\bf P}$, we can determine the system to which users of class $n$ will connect by the value $P_{nl}$.
As an example, knowing the function $f(.)$ and the policy ${\bf P}$, if the network is in state ${\bf M}$, a class $n$ user will connect to system $P_{nl}$,
where $l=f({\bf M})$.

There are some important remarks to keep in mind when speaking about
policies. The first is that we suppose that a user connected to a
system stays within it until the end of his communication in order
to avoid vertical handovers and their signaling overhead.
Furthermore, even if the decision is distributed, all users will
have the same policy and learn together how to enhance it. However,
a policy change will occur after an observation time, long enough to
insure that the steady state of the network has been reached. Note
also that users can connect to a system only if there is room in it,
otherwise they are directed by the network to an available system or
blocked if all systems are saturated.

\section{Utilities}

We analyze a system offering streaming calls.
The goal of a streaming user is to achieve the best throughput, knowing that the different codecs allow a throughput between an upper (best) $T_{max}$ and a lower (minimal) $T_{min}$ bounds. His utility is thus expressed by the quality of the streaming flow he receives, which is in turn closely related to his throughput. Indeed, a streaming call with a higher throughput will use a better codec offering a better video quality. This throughput depends not only on the peak throughput, but also on the evolution of the number of calls in the system where the user decides to connect.
Note that a user that cannot be offered this minimal throughput in neither of the available systems is blocked in order to preserve the overall network performance.

\subsection{Steady state analysis}
\subsubsection{Instantaneous throughput}
The instantaneous throughput obtained by a user in a system depends on the state of the system. The throughput
of a user with radio condition class $n$ connected to system $s$ is given by: \beqq
t_n^s({\bf M})= \min \left[D_n^s
\frac{G({\bf M})}{\sum_{m=1}^N \sum_{r=1}^S M_m^r}, T_{max}\right] \eeqq where
$G({\bf M})$ is the scheduler gain. Note here that the admission control will insure that $t_n^s({\bf M})\geq T_{min}$ by blocking new arrivals. The space of feasible states $\mathcal{A}$ is thus the set of all states ${\bf M}$ where this constraint is ensured:
\begin{equation}
\frac{\sum_{m=1}^N \sum_{r=1}^S M_m^r}{G({\bf M})}\leq \frac{T_{min}}{D_n^s}, \forall n,s|M_n^s>0
\end{equation}

\subsubsection{Steady state probabilities}
The throughput achieved by a user depends on the number of ongoing calls. This latter is a random variable whose evolution is governed by the arrival and departure processes.
We assume that the arrival process of new connections with radio condition $n$ is Poisson with rate
$\lambda_n$. Each arriving user makes a streaming connection whose duration is exponentially distributed with  parameter $1/\mu$.

Within the space of feasible states $\mathcal{A}$, transitions are due to:
\begin{itemize}
\item Arrivals of users of radio condition $n$. Let ${\cal G}_n^s({\bf M})$ denote the state of  the system if we add one
mobile of radio conditions $n$ to system $s$: ${\cal G}_n^H({\bf M})  = (M_1^1 , ... , M_N^1 , ... ,M_1^s,...,M_n^s+1,...,
M_N^s,... , M_1^S , ... , M_N^S )$. The transition from state ${\bf M}$ to ${\cal G}_n^s({\bf M})$ happens if the policy implies that system $s$ is to be chosen for the load information corresponding to state ${\bf M}$, and if the state ${\cal G}_n^s({\bf M})$ is an admissible state. The corresponding transition rate is thus equal to:
\begin{equation}
q({\bf M}, {\cal G}_n^s({\bf M})|{\bf P})=\lambda_n \cdot I_{P_{n,f({\bf M})}=s}\cdot I_{{\cal G}_n^s({\bf M})\in \mathcal{A}}
\end{equation}
where $I_{C}$ is the indicator function equal to 1 if condition $C$ is satisfied and to 0 otherwise.
\item Departures of users of radio condition $n$. Let ${\cal D}_n^s({\bf M})$ denote the state with one less mobile of class $(n,s)$. The transition from state ${\bf M}$ to ${\cal D}_n^s({\bf M})$ is equal to:
\begin{equation}\label{dep}
q({\bf M},{\cal D}_n^s({\bf M})|{\bf P})=M_n^s \cdot \mu \cdot I_{M_n^s>0}
\end{equation}

\end{itemize}
The transition matrix
$\textbf{Q}({\bf P})$ of the Markov process is written for each policy ${\bf P}$ knowing that its diagonal element is:
\begin{equation}\label{qmm}
q({\bf M}, {\bf M}|{\bf P})=-\sum_{n=1}^N \sum_{s=1}^S (q({\bf M}, {\cal D}_n^s({\bf M})|{\bf P})+q({\bf M}, {\cal G}_n^s({\bf M})|{\bf P}))
\end{equation}
The steady-state distribution is then obtained by
solving:
\begin{equation} \left\{
\begin{array}{l}
{\bf \Pi}({\bf P}) \cdot \textbf{Q}({\bf P}) = 0\\
{\bf \Pi}({\bf P}) \cdot {\bf e} = 1;
\end{array}\right.
\label{Q}\end{equation} ${\bf \Pi({\bf P})}$ being the vector
of the steady-state probabilities $\pi({\bf M}|{\bf P})$ under policy ${\bf P}$
and ${\bf e}$ is a vector of ones. 

Once the vector ${\bf \Pi}$ is obtained, the global performance
indicators can be calculated, e.g., the blocking rate of class-$n$ calls knowing that the load information is equal to $l$:
\begin{equation}
b_n(l|{\bf P})=\frac{\sum_{{\bf M}\in \mathcal{A}; {\cal G}_n^s({\bf M})\notin \mathcal{A}, \forall s\in[1,S]}\pi({\bf M}|{\bf P})}{\sum_{{\bf M}\in \mathcal{A}; f({\bf M})=l}\pi({\bf M}|{\bf P})}
\end{equation}
In this equation, we consider as blocked all calls that arrive in states where both systems are saturated, i.e., where $t_n^s({\bf M})<T_{min}$, $\forall j\in {H,L}$. We also obtain the overall blocking rate:
\begin{equation}\label{blockoverall}
b({\bf P})=\sum_{n=1}^N\frac{\lambda_n}{\sum_{m=1}^N\lambda_m}\sum_{{\bf M}\in \mathcal{A}; {\cal G}_n^s({\bf M})\notin \mathcal{A}, \forall s\in[1,S]}\pi({\bf M}|{\bf P})
\end{equation}

\subsection{Transient analysis}
The steady-state analysis described above is not sufficient to
describe the utility of the users as the throughput obtained by a
user at his arrival is not a sufficient indication about the quality
of his communication because of the dynamics of arrivals/departures. In order to obtain the utility, we modify the Markov
chain in order to allow tracking mobiles during their connection
time. For users of radio condition $n$ connected to system $s$, only states where there is at least one user $(n,s)$ are considered. The calculation is as follows:
\begin{enumerate}
\item   Introduce absorbing states $A_n^s$ corresponding to the departure of mobiles that have terminated their connections. Additional transitions are thus added between ${\bf M}$  and $A_n^s$ with rate equal to:
\begin{equation}
\tilde{q}_n^s({\bf M},A_n^s)=\mu \cdot I_{M_n^s>0}
\end{equation}
The transitions to the neighboring states with one less user are then modified accordingly by subtracting $\mu$ from the original transition rates defined in equation (\ref{dep}):
\begin{equation}
\tilde{q}_n^s({\bf M},{\cal D}_n^s({\bf M})|{\bf P})=(M_n^s-1) \cdot \mu \cdot I_{M_n^s>0}
\end{equation}
The remaining transition rates remain equal to the original transitions:
\begin{displaymath}
\tilde{q}_n^s({\bf M},{\cal G}_{n^\prime}^{s^\prime}({\bf M})|{\bf P})=q({\bf M},{\cal G}_{n^\prime}^{s^\prime}({\bf M})|{\bf P}),\quad \forall n^\prime ,s^\prime 
\end{displaymath}
and
\begin{displaymath}
\tilde{q}_n^s({\bf M},{\cal D}_{n^\prime}^{s^\prime}({\bf M})|{\bf P})=q({\bf M},{\cal D}_{n^\prime}^{s^\prime}({\bf M})|{\bf P}),\quad \forall n^\prime ,s^\prime\neq s 
\end{displaymath}
\item Define matrix $\tilde{{\bf Q}}_n^s$ of elements $\tilde{q}_n^s({\bf M},{\bf M}^\prime)$ defined above and with diagonal elements as in equation (\ref{qmm}):
\begin{displaymath}
\tilde{q}_n^s({\bf M},{\bf M}|{\bf P})=q({\bf M},{\bf M}|{\bf P})
\end{displaymath}
Under policy ${\bf P}$ , the volume of information $I_n^s({\bf M}|{\bf P})$ sent by system $s$ users subject to radio conditions $n$ starting from state ${\bf M}$ is then equal to the volume of information sent between state ${\bf M}$ and the absorbing state $A_n^s$.  These values can be calculated by solving the set of linear equations  for all states ${\bf M}$:
\begin{equation}\label{volume}
\sum \tilde{q}_n^s({\bf M},{\bf M}^\prime|{\bf P})I_n^s({\bf M}^\prime|{\bf P})=-t_n^s({\bf M})
\end{equation}
knowing that $I_n^s(A_n^s)=0$.
\item The utility of a class-$n$ user that has
found the network in state ${\bf M}$ and chosen to connect to system $s$ is the volume of information sent starting from state ${\cal G}_n^s({\bf M})$. Recall that ${\cal G}_n^s({\bf M})$ is defined as the state with one more class-$n$ call connected to system $s$:
\begin{equation}
u_n^s({\bf M}|{\bf P})=I_n^s({\cal G}_n^s({\bf M})|{\bf P})
\end{equation}
\end{enumerate}

\section{Optimality, game and control}
In this section, we use the utilities of users that we obtained above to derive the association policies. We first search for the optimal policy, i.e. the policy that maximizes the global utility of the network. Nevertheless, as it is not realistic to consider that the users will seek the global optimum, we show how to find the policy that corresponds to the Nash equilibrium, knowing that users will try to maximize their individual utility. We will next show how the operator can control, by sending appropriate load information, the equilibrium of its wireless users to maximize its own utility.

\subsection{Optimality}
\subsubsection{Global utility}
When a global optimum is sought, it is important to maximize the QoS of all users. The global utility function can be written as:
\begin{eqnarray}\label{Global_Utility}
U({\bf P})=\sum_{n=1}^N\frac{\lambda_n}{\sum_{i=1}^N \lambda_i}\sum_{l\in \mathfrak{L}}[(1-b_n(l|{\bf P}))\times\\\nonumber\sum_{{\bf M}|f({\bf M})=l}u_n^{(P_{n,l})}({\bf M}|{\bf P})\pi({\bf M}|{\bf P})]
\end{eqnarray}
knowing that $P_{n,l}\in[1,S]$ is the system where new users of class-$n$ connect when they receive the load information $l$ and have the policy ${\bf P}$.

Note that, in this utility, we consider not only the QoS of accepted users (throughput), but also the blocking rate as the aim is also to maximize the number of accepted users. We also weight the users with different radio conditions with their relative arrival rates.

\subsubsection{Optimal policy}
Knowing the utility in equation (\ref{Global_Utility}), the optimal policy is the one among all possible policies that maximizes this utility:
\begin{equation}
{\bf P}^*=\argmax{{\bf P}}U({\bf P})
\end{equation}


\subsection{Equilibrium}
\subsubsection{Individual utility}

If the aim is to maximize the individual utility, users of different
radio conditions are interested by maximizing the QoS they
obtain given the load information broadcast by the network. The utility that a class $n$ user might obtain if he chooses system $s$ when the load information is $l$, while all other users follow policy ${\bf P}$ is then:
\begin{equation}\label{Individual_Utility}
U_{nl}^s({\bf P})=\frac{\sum_{{\bf M}|f({\bf M})=l}u_n^{s}({\bf M}|{\bf P})\pi({\bf M}|{\bf P})}{\sum_{{\bf M}|f({\bf M})=l}\pi({\bf M}|{\bf P})}
\end{equation}

\subsubsection{Nash equilibrium}

A policy ${\bf P}^*$ corresponds to a Nash equilibrium if, for all
radio conditions and all load information, the individual utility
obtained when following ${\bf P}^*$ is the largest possible utility
under ${\bf P}^*$. Mathematically, this can be expressed by the
following inequality for all radio conditions $n\in [1,N]$ and all
load information $l\in[1,L]$:

\begin{equation}
U_{nl}^{(P^*_{n,l})}({\bf P}^*)\geq U_{nl}^s({\bf P}^*),\forall s\in[1,S]
\end{equation}

\subsection{Control}

In the previous section, we derived the policy that corresponds to
the Nash equilibrium for a game where players are the wireless users
that aim at maximizing their utility. However, there is another
dimension of the problem related to the information sent by the
network and corresponding to the different load information.
Motivated by the fact that the network may guide users to an
equilibrium that optimizes its own utility if he chooses the
adequate information to send, we introduce a control problem, that can also be modeled as a game between the base station and its users. At the
core lies the idea that introducing a certain degree of hierarchy
in non-cooperative games not only improves the individual efficiency
of all the users but can also be a way of reaching a desired trade-off between
the global network performance at the equilibrium and the requested
amount of signaling. 

More formally, the way of aggregating the loads in the broadcast information (expressed by the function $f(.)$) is inherent to the previous analysis. In particular, the utilities of individual users, calculated in equation (\ref{Individual_Utility}), is function of $f(.)$:
\begin{displaymath}
U_{nl}^s({\bf P}|f)=\frac{\sum_{{\bf M}|f({\bf M})=l}u_n^{s}({\bf M}|{\bf P})\pi({\bf M}|{\bf P})}{\sum_{{\bf M}|f({\bf M})=l}\pi({\bf M}|{\bf P})}
\end{displaymath}

This leads the wireless users to a Nash equilibrium that depends on the way the network aggregates the load information:
\begin{displaymath}
{\bf P}^*={\bf P}^*(f)
\end{displaymath}

The control problem is thus defined as the maximization of the utility of the network by tuning the function $f(.)$. If the aim of the operator is to maximize its revenues by maximizing the acceptance ratio, the optimal solution is:
\begin{equation}
f^*=\argmax{f}\frac{1}{b({\bf P}^*(f))}
\end{equation}
with blocking defined as in equation (\ref{blockoverall}).

\section{Results}
For illustration, we consider the case of a network composed of
HSDPA and 3G LTE systems. Users are classified between users with
good radio conditions (or cell center users) and users with bad
radio conditions (or cell edge users). The network sends aggregated
load information as shown in Figure \ref{fig:pload} with the
following thresholds: $[H_1=0.3,H_2=0.7,L_1=0.3,L_2=0.7]$, meaning
that a system is considered as highly loaded if its load exceeds
$0.7$ and as low-loaded if its load is below $0.3$.

We also consider a streaming service where users require a minimal throughput of 1
Mbps and can profit from throughputs up to 2 Mbps in order to enhance video quality ($D_{min}=1 Mbps$ and $D_{max}=2 Mbps$). We consider an offered traffic that varies from
1 to 10 Erlangs and obtain numerically the equilibrium points.

\subsection{Equilibrium}

We focus on the Nash equilibrium when wireless users aim at maximizing their individual utility. For comparison purposes, we study three different association approaches:
\begin{itemize}
\item Hybrid decision approach: The proposed hybrid scheme where users receive aggregated load information and aim at maximizing their individual utility. We illustrate the global utility corresponding to the Nash equilibrium policy.
\item Peak rate maximization approach: This is a simple association scheme where users do not have any information about the load of the systems. They connect to the system offering them the best peak rate:
\begin{displaymath}
s^*= \argmax{s}D_n^s \end{displaymath}
Note that this peak rate can be known by measuring the quality of the receiving signal.
\item Instantaneous rate maximization approach: The network broadcasts ${\bf M}$, the exact numbers of connected users with different radio conditions. Based on this information and on the measured signal strength, the wireless users estimate the throughput they will obtain in both systems.  A new user with radio condition $n$ will then connect to the system $s^*$ offering him the best throughput:
\begin{displaymath}
s^*= \argmax{s}\frac{D_n^s}{1+\sum_{m=1}^N \sum_{r=1}^S M_m^r} \end{displaymath}
Note that this scheme is not realistic as the network operator will not divulge the exact number of connected users in each system and each position of the cell.
\end{itemize}

\begin{figure}[ht]
\centering
\includegraphics[width=8cm]{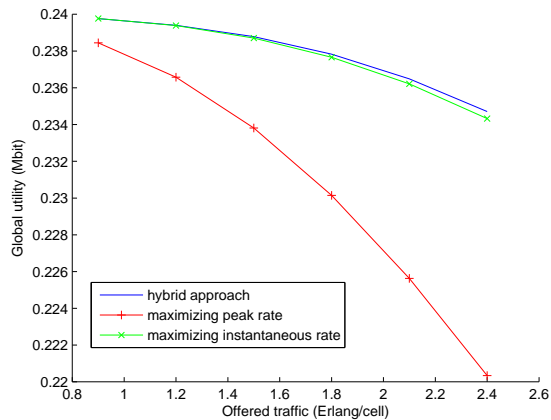}
\caption{Global utility.} \label{utility}
\end{figure}

We plot in Figure \ref{utility} the global utility for the three cases. This global utility is the one defined in equation (\ref{Global_Utility}) and expressed in
Mbits, as users are
interested in maximizing the information they send during their
transfer time.

As intuition would expect, the results show that the peak rate
maximization approach has the worst performance as a system that
offers the largest peak throughput may be highly-loaded, resulting
in a bad QoS. However, a surprising result is that the hybrid
scheme, based on partial information, is comparable and even
outperforms the full information scheme when traffic increases. This
is due to the fact that streaming users will have relatively long
sessions, visiting thus a large number of network states; knowing
the instantaneous throughput at arrival will not bring complete
information about the QoS during the whole connection. On the
contrary, the proposed hybrid, game theoretic, approach aims at
maximizing the QoS during the connection time.

\subsection{Control}
\begin{figure}[ht]
\centering
\includegraphics[width=8cm]{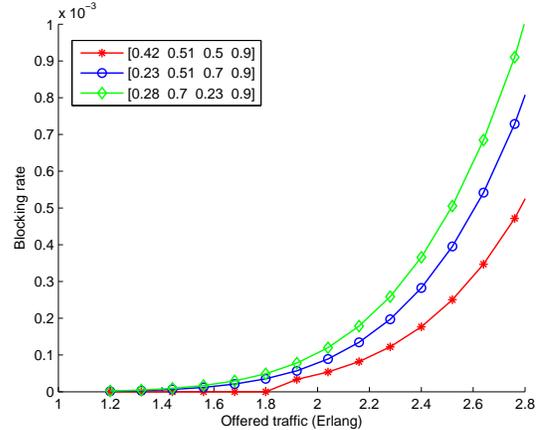}
\caption{Blocking rate for different broadcast load information; a
vector of thresholds $[H_1,H_2,L_1,L_2]$ means that system $s$ will
be considered as highly loaded if its load exceeds $s_2$ and as
low-loaded if its load is below $s_1$ ($s=H$ for HSDPA and $L$ for
LTE).} \label{hier}
\end{figure}

We now turn to the second stage of our problem, where the network
tries to control the users' behavior by broadcasting appropriate
information, expected to maximize its utility while individual users
maximize their own utility. We plot in Figure \ref{hier} the
blocking rate for different ways of aggregating load information,
obtained when users follow the policy corresponding to Nash
equilibrium. In this figure, we plot the results for three cases:
the optimal thresholds (in red stars) and two other sets of
thresholds. We can observe that the utility of the network
(expressed in the acceptance rate) varies significantly depending on
the load information that is broadcast. Such an accurate modeling of
the control problem is a key to understand the actual benefits
brought by the proposed hybrid decision approach.

\section{Conclusion}
In this paper, we studied hybrid association schemes in
heterogeneous networks. By hybrid schemes we mean distributed
decision schemes assisted by the network, where the wireless users
aim at maximizing their own utility, guided by information broadcast
by the network about the load of each system. We first show how to
derive the utilities of flows that are related
to the QoS they receive under the different association policies. We
then derive the policy that corresponds to the Nash equilibrium. Finally, we show how the operator, by sending appropriate information about the state of the
network, can optimize its own utility. The proposed hybrid decision approach for the association
problem can reach a good
trade-off between the global network performance at the equilibrium
and the requested amount of signaling. 
\section*{Acknowledgment}
This work was supported by the ANR project WiNEM.

\end{document}